\documentclass{article}

\usepackage{PRIMEarxiv}
\usepackage{subcaption}
\usepackage[utf8]{inputenc} 
\usepackage[T1]{fontenc}    
\usepackage{hyperref}       
\usepackage{url}            
\usepackage{booktabs}       
\usepackage{amsfonts}       
\usepackage{nicefrac}       
\usepackage{microtype}      
\usepackage{lipsum}
\usepackage{fancyhdr}       
\usepackage{graphicx}       
\graphicspath{{media/}}     

\title{Searching Clinical Data using Generative AI
}

\author{
  Karan Hanswadkar,    Anika Kanchi,    Shivani Tripathi,     Shi Qiao,     Rony Chatterjee,     Alekh Jindal \\
  Tursio Inc., USA \\
}

\begin{document}
\maketitle

\begin{abstract}
  Artificial Intelligence (AI) is making a major impact on healthcare, particularly through its application in natural language processing (NLP) and predictive analytics. The healthcare sector has increasingly adopted AI for tasks such as clinical data analysis and medical code assignment. However, searching for clinical information in large and often unorganized datasets remains a manual and error-prone process. Assisting this process with automations can help physicians improve their operational productivity significantly. 
  
  In this paper, we present a generative AI approach, coined {\it SearchAI}, to enhance the accuracy and efficiency of searching clinical data. Unlike traditional code assignment, which is a one-to-one problem, clinical data search is a one-to-many problem, i.e., a given search query can map to a family of codes. Healthcare professionals typically search for groups of related diseases, drugs, or conditions that map to many codes, and therefore, they need search tools that can handle keyword synonyms, semantic variants, and broad open-ended queries. SearchAI employs a hierarchical model that respects the coding hierarchy and improves the traversal of relationships from parent to child nodes. SearchAI navigates these hierarchies predictively and ensures that all paths are reachable without losing any relevant nodes. 
    
  To evaluate the effectiveness of SearchAI, we conducted a series of experiments using both public and production datasets. Our results show that SearchAI outperforms default hierarchical traversals across several metrics, including accuracy, robustness, performance, and scalability. SearchAI can help make clinical data more accessible, leading to streamlined workflows, reduced administrative burden, and enhanced coding and diagnostic accuracy. 
  \end{abstract}
  

\section{Introduction}
%
%
%
%

Generative AI is quickly becoming an integral part of modern healthcare due to its capabilities in interpreting and analyzing natural language. No wonder the AI healthcare market was valued at USD $\$11$ billion in 2021 and is projected to reach USD $\$187$ billion by 2030~\cite{ai-stats}. These are massive projections, suggesting significant changes in how medical providers, hospitals, pharmaceuticals, biotech companies, and others in the healthcare industry operate. Interestingly, clinical data is a common thread amongst these different stakeholders, with applications such as identifying patients for clinical trials, analyzing trends for medication and lab tests, and monitoring patient outcomes. Unfortunately, clinical data is challenging since it is often incomplete or messy, making it difficult for accurate analyses.

To illustrate, diagnosis is a challenge in clinical data and leads to some of the most common medical errors. Researchers at the University of California, San Francisco, conducted a study involving seriously ill patients from academic medical centers across the country~\cite{uc-study,10.1001/jamainternmed.2023.7347}. Their findings revealed that nearly a quarter of these patients had experienced delayed diagnoses. Such errors led the patients to be transferred to ICUs and even played a role in about one in fifteen deaths~\cite{10.1001/jamainternmed.2023.7347}. The consequences could be severe, ranging from chronic to long-term health problems for patients. The researchers further estimated that eliminating assessment and testing problems could reduce the risk of diagnostic errors by approximately $40\%$. Reducing diagnostic errors can also result in up to $50\%$ lower treatment costs and $40\%$ better health outcomes, apart from improved monitoring and preventive care~\cite{ibm-healthcare-benefits}.

The traditional approach to making clinical data useful is to standardize various pieces of information using a common coding ontology, e.g., ICD codes for diseases~\cite{icd-codes}, LOINC codes for lab tests~\cite{loinc-codes}, and CPT codes for drug administration~\cite{cpt-codes}. 
Historically, these codes were assigned manually, but newer AI tools assist physicians in coding the clinical data at the source. For example, Spark NLP from Job Snow Labs assigns ICD-10 codes with $76\%$ accuracy~\cite{spark-nlp} and Miraico from ASUS makes AI-based recommendations with $99.4\%$ accuracy~\cite{miraico}. However, searching clinical data that contains these codes is still an open problem.

In this paper, we describe a generative AI approach to searching clinical data.
Our goal is to help healthcare professionals, including physicians, clinical coders, and researchers, to efficiently retrieve the relevant pieces of data. 
For example, a physician can search for ``fever and cough" and the system would retrieve relevant ICD codes from the dataset that are related to respiratory infections, guiding the physician towards accurate diagnosis and faster decisions.
However, the key challenge here is to map the input query to the broader criterion of diseases or drugs codes that are relevant. This contrasts with the code assignment problem, which is a one-to-one problem since a given disease or drug can only be mapped to one code. Furthermore, given that the coding ontologies are hierarchical, the parent-child relationships must be obeyed when generating the criterion --- a tedious task since the hierarchies were originally designed for manual use by humans and automating them programmatically requires further tuning for effective disambiguation. 



\begin{table*}[!t]
\centering
\small
\begin{tabular}{|c|l|l|}
\hline
Chapter & Code & Range Description\\
\hline
1 & A00-B99 & Certain Infectious and Parasitic Diseases\\
2 & C00-D49 & Neoplasms\\
3 & D50-D89 & Diseases of the Blood and Blood-Forming Organs and Certain Disorders Involving\\
& &  the Immune Mechanism\\
4 & E00-E89 & Endocrine, Nutritional and Metabolic Diseases\\
5 & F01-F99 & Mental, Behavioral and Neurodevelopmental Disorders\\
6 & G00-G99 & Diseases of the Nervous System\\
7 & H00-H59 & Diseases of the Eye and Adnexa\\
8 & H60-H95 & Diseases of the Ear and Mastoid Process\\
9 & 100-199 & Diseases of the Circulatory System\\
10 & J00-J99 & Diseases of the Respiratory System\\
11 & K00-K95 & Diseases of the Digestive System\\
12 & L00-L99 & Diseases of the Skin and Subcutaneous Tissue\\
13 & M00-M99 & Diseases of the Musculoskeletal System and Connective Tissue\\
14 & N00-N99 & Diseases of the Genitourinary System\\
15 & 000-09A & Pregnancy, Childbirth and the Puerperium\\
16 & P00-P96 & Certain Conditions Originating in the Perinatal Period\\
\hline
\end{tabular}
\caption{Snapshot of ICD-10-CM chapters and code ranges.}
\label{tab:icd-snapshot}
\vspace{-0.2cm}
\end{table*}

Our key ideas are three-fold. First, we train small models to decompose patient search queries into the underlying Boolean logic. Second, we train a hierarchical model to traverse the ontology while respecting parent-child relationships. And finally, we tune the hierarchies for specific database instances to narrow down the scope and make the patient search tailored. 
Unlike a naive SQL query generator, SearchAI leverages natural language processing (NLP) techniques to tokenize query text into meaningful units, and identify medical entities for diseases, symptoms, and procedures. 
The goal is to reduce the time and effort involved in searching clinical data,
allowing healthcare professionals to focus more on patient care and less on administrative tasks. 
In summary, our key contributions are as follows:

\begin{enumerate}
\item We describe the clinical data search problem and how it differs from traditional code assignment. (Section~\ref{sec:overview})
\item We introduce SearchAI, a generative AI-based approach for searching clinical data. SearchAI decomposes natural language questions into one or more Boolean predicates and translates each predicate to one or more code comparison in the underlying database. (Section~\ref{sec:searchai})
\item We present a hierarchical algorithm for accurately mapping filter predicates to the family of relevant codes. The algorithm traverses the code hierarchy predictively while ensuring all paths are reachable. (Section~\ref{sec:hierarchical-predictor})
\item Finally, we present an evaluation using both public and production datasets to evaluate the accuracy, robustness, performance, and scalability of SearchAI. (Section~\ref{sec:experiments})
\end{enumerate}


\section{Overview}
\label{sec:overview}

One of the key challenges in the healthcare domain is to retrieve relevant information using the coding system. The International Classification of Diseases (ICD) is a medical coding system designed by the World Health Organization (WHO). It dates to the 1850s and is still used to catalog health conditions and map complex diseases. As modern healthcare evolved, new iterations of ICD codes have been introduced over the years. The latest iteration, ICD-10, significantly improves upon previous versions by increasing capacity and allowing for better indexing and organization. Table~\ref{tab:icd-snapshot} shows an example of how these ICD codes are organized based on their code ranges.
For instance, code range C00-D49 is for neoplasms, and code range J00-J99 is for the respiratory system.

Unfortunately, despite all the progress in healthcare, as of 2024, ICD code lookup remains largely a manual process. This manual coding relies on professionals with specialized skills such as medical knowledge, familiarity with coding regulations, and expertise in clinical terminology~\cite{sciencedirect}. This approach is not only time-consuming and labor-intensive but also susceptible to human errors and inconsistencies across coders. “For one hospital admission, it takes half an hour for manual coding,” and in the United States, “the cost of coding mistakes is estimated at \$ 25 billion annually”~\cite{sciencedirect}. Moreover, coders require continuous training to stay current with new ICD versions; for instance, ICD-10 “offers a more extensive and detailed classification system,” which “demonstrates enhanced specificity with 23,000 six-character codes—twice the number found in ICD-9”~\cite{sciencedirect}. Given these limitations, automatic coding has increasingly attracted the attention of researchers.

\subsection{Complexity of the ICD Codes}

\begin{figure}[!t]
\centering
\includegraphics[width=0.45\textwidth]{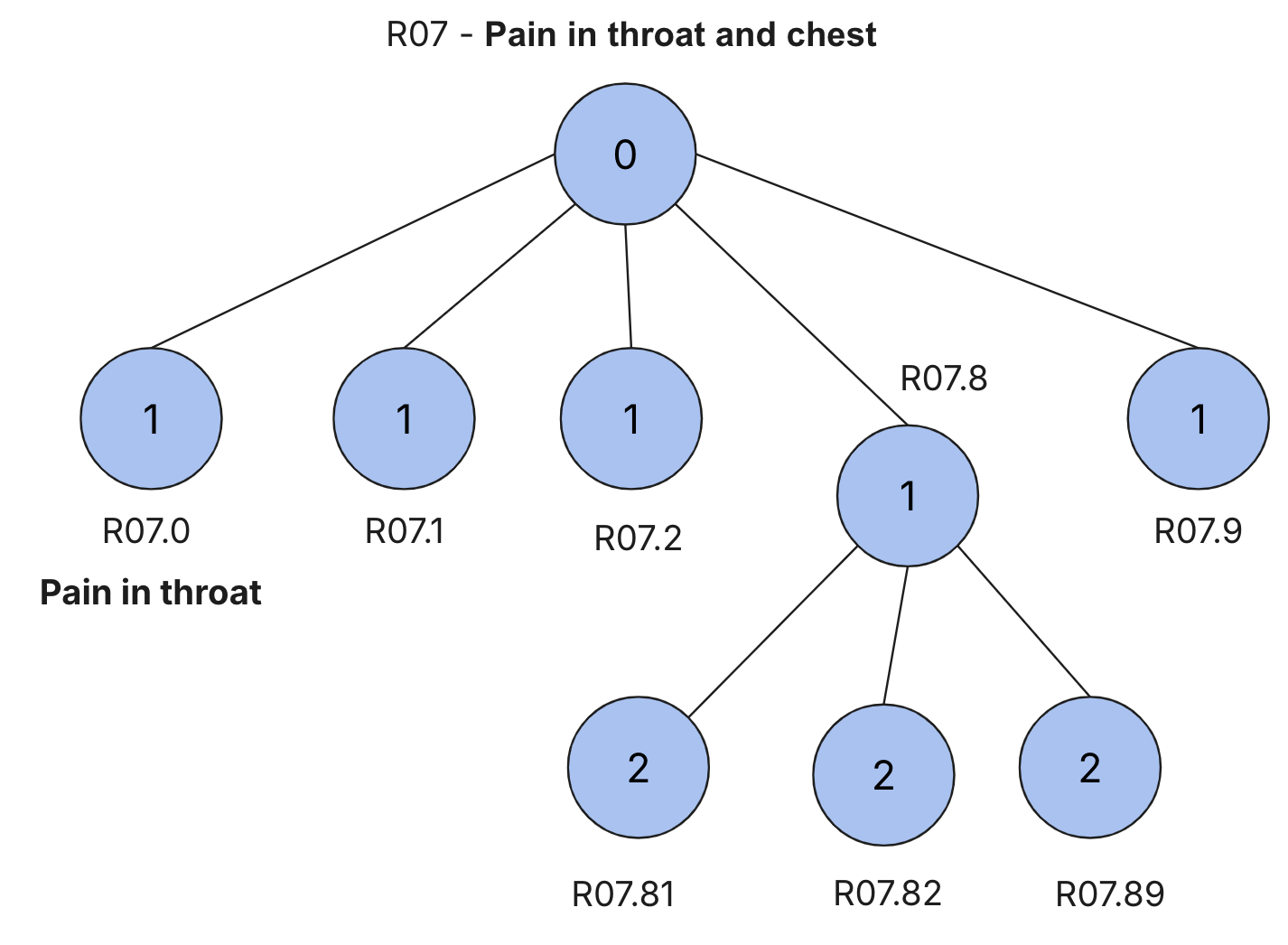}
\caption{Hierarchical structure of ICD-10 code R07 (Pain in throat and chest).}
\label{fig:icd-hierarchy}
\end{figure}

Figure~\ref{fig:icd-hierarchy} shows ICD code R07, representing the diagnosis of ``Pain in the throat and chest''. We can see that R07 is just one of the many entries that can be found in the list of ICD-10 codes, and we can visualize them in a tree structure. The root of the tree, representing the level-0 ICD code, corresponds to the broader category of diseases or health conditions. The descendants of the root, i.e., ICD codes at levels 1, 2, and 3, are identified by their decimal extensions and are typically grouped based on factors such as location, severity, or specific condition names. As we go deeper down the tree, the categories become increasingly more specific and detailed. For example, R07.0 (``Pain in throat'') is more specific compared to R07 (``Pain in the throat and chest'').
 
Analyzing the ICD-10 data is complex due to several reasons:
(i)~large-sized hierarchies with many descendant codes under a single root code, making it difficult to capture the full breadth of relationships,
(ii)~ambiguous description of closely related ICD codes, making it hard to interpret and classify, 
(iii)~inconsistent pattern of parent-child descriptions, which can either narrow down or choose distinct terminology to make the disambiguation harder.
In this paper, our goal is to address these issues for a given instance of the code dataset. We want to refine the code data and optimize it for algorithmic traversal. Additionally, 
we want to ensure that each node in the hierarchy remains accessible.

\subsection{Challenges of the problem} 

ICD codes have overlapping descriptions that can be difficult to interpret with any automation. For instance, many root-level code descriptions contain terms such as ``and'' or ``with'', as seen earlier in code description for R07 (``Pain in throat and chest.''). The problem with these descriptions is that if we were to search for a descendant like R07.0 ( ``Pain in the throat''), the root code would also be considered, since the exact text tokens appear in both parent and child.
Having both parent and child codes are helpful for a broader search but not for a specific one.
The flip side of returning strict matches is false negatives, i.e., missing codes that users might expect.
In our current accuracy tests, when a system returns multiple ICD codes, we do not automatically label it as incorrect. Instead, we carefully review and assess whether the result produced is truly incorrect. Examining the occurrences of false negatives gives a better understanding of the ICD lookup accuracy.

Furthermore, handling commonly seen vague terms such as ``unspecified'', ``other'', and ``classified'' has proven to be difficult. Such words appear frequently across various levels, making an automated approach to consider them as important and misleading into multiple different paths. Consequently, the result could be incorrect or completely disparate set of ICD codes.
Therefore, we need better strategies to handle vague terms without compromising the overall accuracy of the code search process.

Finally, flattening the hierarchy could be one way to simplify the search process. However, this is very risky: while its likely to improve accuracy at deeper levels (e.g., levels 2 and 3), the accuracies closer to the root level (e.g., levels 0 and 1) get worse. We saw this experimentally in our approach, where flattening the hierarchies in levels 1, 2 and 3, had reduced the accuracy at level-0 from $97\%$ to approximately $85\%$.




\section{SearchAI}
\label{sec:searchai}

\begin{figure*}[!t]
    \centering
    \includegraphics[width=0.9\textwidth]{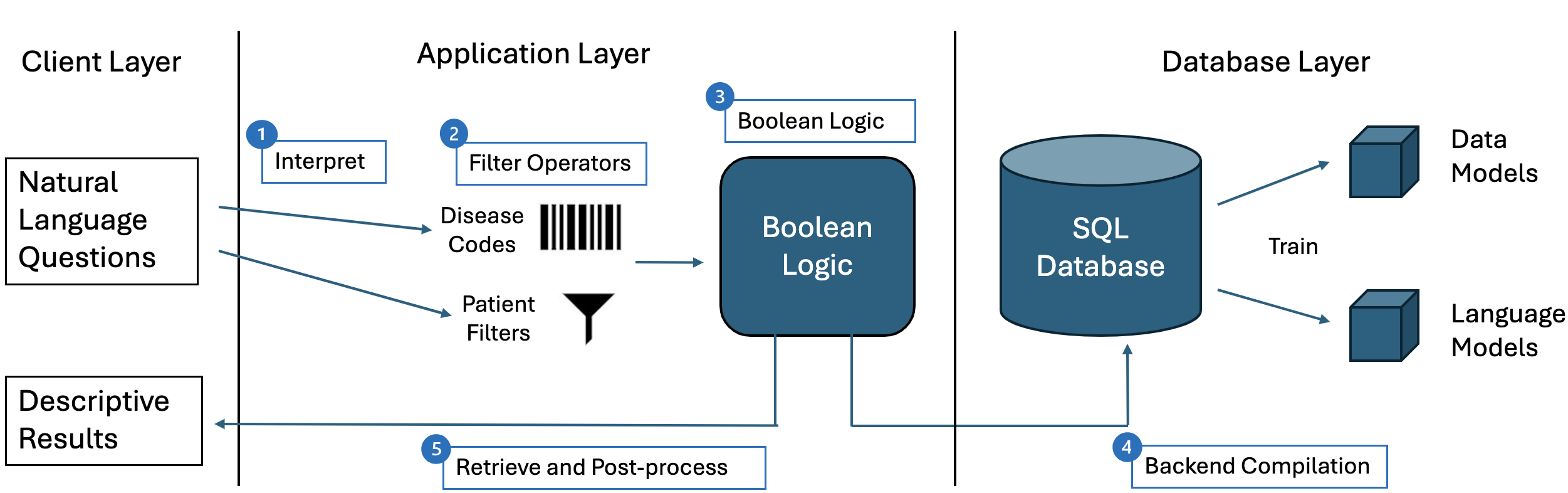}
    \vspace{0.2cm}
    \caption{The model architecture of SearchAI.}
    \label{fig:search_ai}
    \end{figure*}
    
In this section, we describe the overall design of SearchAI. Figure~\ref{fig:search_ai} shows the architecture, where users provide natural language questions and SearchAI generates accurate descriptive answers. To illustrate, users can ask the following types of questions and expect to see the relevant patient population:

\begin{itemize}
    \small
    \item Show sepsis patients.
    \item Show patients who are diagnosed with Anemia.
    \item Show patients who have external causes of abnormal reaction for surgical operations.
    \item Show patients with external causes of falls on the same level as slipping.
    \item Show patients with chronic ischemic heart disease.
    \item Show patients who are on psychoactive substance use.
    \item Show patients diagnosed with Type 2 diabetes mellitus with diabetic nephropathy.
    \item Show patients diagnosed with prediabetes and undergo drug abuse counseling.
\end{itemize}

Interpreting the above questions and generating responses involves a sequence of five steps. We describe each one of them below.

\paragraph*{1. Interpret.} First, SearchAI parses the user queries into smaller meaningful fragments, e.g., the users might ask for one or more diseases, drugs, or patient demographics. Breaking down user questions into relevant and well-formed fragments is a critical step in interpreting them accurately. This allows users to compose extended complex questions since the system will anyway break them down into smaller understandable fragments. For each fragment, SearchAI interprets the relevant filter predicates. For medical codes, SearchAI uses a hierarchical predictor that obeys the coding hierarchy and generates the set of all applicable codes for each filter predicate. For patient demography or other metadata filters, SearchAI identifies the literals to be narrowed down, e.g., state, age group, family history, and so on.

\paragraph*{2. Filter Operators.} Once the question is interpreted into filter predicates, SearchAI constructs the filter operators based on the underlying database schemas, i.e., which column or sets of columns need to be filtered upon, the data types involved (numeric, string, dates, etc.), and the comparison operation required (range, equality, inequality, similarity, etc.). SearchAI supports both keyword and semantic search for similarity filters. For instance, the severity of the condition might be implied in the diagnosis description, which users can filter semantically.

\paragraph*{3. Boolean Logic.} SearchAI composes filter operators into Boolean logic, i.e., combinations of AND, OR, and NOT. This is a particularly hard step for practitioners since they need to carefully apply the reasoning based on what they are trying to achieve. For instance, getting patients from Washington and Oregon states is an OR logic on the State column, whereas getting patients under 50 in Washington is an AND logic on two separate columns. SearchAI identifies relevant columns from the underlying database and applies correspondingly valid Boolean logic.

\paragraph*{4. Backend Compilation.} SearchAI compiles the resulting Boolean logic against the backend database, i.e., considering the table schemas, key-foreign relationships for searching across multiple tables, and any data models that could be leveraged for query acceleration. The system manages a set of specialized data models that can narrow down the search to specific portions of the data quickly.

\paragraph*{5. Retrieve and Post-process.} SearchAI executes the compiled queries and retrieves the results from the underlying database. It caches partial results to improve performance and keeps end-to-end latencies interactive, i.e., within $3$ seconds for the $95^{th}$ percentile. Once retrieved, we present the results in a descriptive manner, including a short summary of the patient population and disease characteristics, trends of drugs and other administrations, and a table of all columns that were relevant and retrieved.

\vspace{0.2cm}
Overall, SearchAI is a brand-new way of interacting with clinical data in natural language. Users can explore patient or disease cohorts, examine clinical data related to diagnoses and medications, and draw inferences easily. The AI-driven process can help them save valuable time and arrive at better outcomes. The rest of the paper focuses on inferring ICD-10 codes from natural language, given that it is a key operation when searching clinical data.


\section{Hierarchical Predictor}
\label{sec:hierarchical-predictor}

In this section, we describe the hierarchical predictor for inferring ICD-10 codes from natural language. Recall that the key requirement is to respect the parent-child hierarchy and return {\it all} applicable codes for the given question. Below, we first describe the base algorithm and then present more advanced variants, along with their pros and cons. We discuss an experimental evaluation of these different variants in Section~\ref{sec:experiments}.



\subsection{Default Hierarchical Predictor}
 
The base version of the algorithm traverses the code hierarchy, assigns similarity scores, groups the matched codes by their scores, and produces the most likely matches as results. This algorithm uses a {\it match dictionary} to keep track of matched elements and groups them by their match scores. 
The pseudocode below illustrates the default hierarchical predictor algorithm in matching user queries with ICD code descriptions. 


\paragraph*{Pseudocode for Default Hierarchical Predictor}
\begin{enumerate} 
  \item Initialize the Matches Dictionary, which creates an empty dictionary to store matched elements grouped by match scores.
  \item Process and iterate through each description in the ICD-10 code dataset.
  \begin{enumerate}
    \item Removing special characters (e.g., punctuation). 
    \item Converting all text to lowercase for case-insensitive matching. 
    \item Tokenize the cleaned description into small, well-formed tokens. These tokens make it easier to match codes to query fragments. 
  \end{enumerate}
  \item Check for exact matches by comparing the text tokens from the query fragment descriptions against the tokens from each ICD-10 code description. 
  \begin {enumerate}
  \item If there are any exact matches (i.e., the tokens appear in both the user's input query and the dataset description), record these matches in the matches' dictionary. 
  \item Group the matches by the number of matching tokens, which serves as their match score. 
  \end{enumerate}
  \item If no exact matches were found during the initial exact match comparison, the algorithm moves on to approximate matching. 
  \begin {enumerate}
  \item Calculate the similarity between the cleaned dataset description and the user input using the Levenshtein distance function for textual similarity and cosine distance between embeddings for semantic similarity.  
  \item If the similarity exceeds the defined threshold, the description is considered a potential match. 
  \end{enumerate}
    \item Once all ICD-10 codes have been processed and the matches are collected, we identify the highest match score and return both the score and the corresponding matches (ICD-10 codes and descriptions). In case of no matches, we indicate that no results were found for the user query.
\end{enumerate}

The above default predictor shows promising results at the root level (level-0), achieving an impressive 97\% accuracy. However, accuracy drops significantly to 57\% for level-1 codes, highlighting some of the critical areas for improvement. Errors included ICD codes being labeled as not found, incorrect matches to different ICD codes, or multiple ICD codes being assigned to a single code. 
The primary cause of inaccuracies in level-1 hierarchies stems from discrepancies in syntax between the level-0 root descriptions and the more specific level-1 descriptions. These differences caused SearchAI to skip relevant search paths and thus miss out on the accurate matches.  


To address the above issues, we extended the default hierarchical predictor by applying a token promotion strategy. The idea is to promote unique keywords from level-1 descriptions into the level-0 root descriptions. This approach improved accuracy for results previously labeled as not found while preserving the hierarchical structure. However, despite these advancements, we were unable to achieve a substantial improvement in accuracy. Turns out that the bigger problem is caused by vague terms like ``Unspecified'', ``Other'', and ``Not classified elsewhere'', which appear frequently in ICD code descriptions at all hierarchy levels. An obvious fix could be to assign lower weights to ambiguous terms, thereby deprioritizing them during the match process. Unfortunately, this only led to moderate improvements, which were insufficient to meet the desired accuracy goals.

\subsection{Hierarchy Flattening}

The code hierarchies were originally designed for human consumption, i.e., physicians, nurses, or research coordinators would sift through them manually and pick the right codes. Making those hierarchies machine programmable requires them to be tuned such that all nodes are reachable. Therefore, we extended the default hierarchical predictor by optimizing the hierarchical structure in the first place. Specifically, we compare the text tokens in child and parent descriptions to identify unmatched entries, i.e., a child that could not be reached. In such a case, we flatten the hierarchy by recursively adjusting the depth levels. While the default algorithm achieved a $57\%$ accuracy, the hierarchy flattening approach further improved the accuracy to $74\%$. We describe the algorithm below.


\paragraph*{Pseudocode for Hierarchical Flattening}
\begin{enumerate}
  \item Initialize an empty list to store node the flattening candidates. 
  \begin{enumerate}
    \item Iterate over each entry in the ICD-10 code dataset 
    \item For Entries at depth 1 
    \begin{enumerate}
      \item Removing special characters (e.g., punctuation).
      \item Converting all text to lowercase for case-insensitive matching.  
      \item Tokenize the cleaned description into individual tokens. These tokens make it easier to perform exact matches. 
    \end{enumerate}
    \item Check for overlaps between child and parent tokens. If there is no overlap, add the entry to the list of flattening candidates that we initialized at the beginning.
  \end{enumerate}
  \item Flatten the level-1 ICD codes by recursively adjusting their depths by 1.
    \begin{enumerate}
      \item For each entry in the list of flattening candidates:
      \begin{enumerate}
        \item Reduce its depth by 1. 
        \item Adjust the depths of its descendants recursively.  
      \end{enumerate}
    \end{enumerate}
    \item Repeat steps 1 and 2 at all depth levels until all elements in the hierarchy are reachable.
    \item Persist the adjusted node entries as the flattened ICD-10 code dataset that is now tuned for more effective search operations algorithmically. 
\end{enumerate}


Our experiments show that the above hierarchy refinement results in better accuracies. 
By flattening the hierarchy where necessary and carefully reviewing incorrect results, we ensured accurate matches and created a scalable solution capable of handling complex ICD-10 code structures. Note that the above process makes SearchAI specialized to the given patient database instead of trying to optimize for all possible codes. This makes sense since data from a given clinic or facility is likely focused on a smaller subset of codes anyways. 


\subsection{Hybrid Approach}

Flattening is good for localizing specific hierarchy levels, but it does not maintain the same accuracy across all levels. 
To handle this, our final variant of the hierarchical predictor first identifies the starting point using a random search before traversing the hierarchies. 

\paragraph*{Pseudocode for Hybrid Hierarchical Predictor}
\begin{enumerate} 
  \item Initialize a Matches Dictionary to store matched elements grouped by their match scores.
  \item Process and iterate through each description in the ICD-10 code dataset.
  \begin{enumerate}
    \item Remove special characters (e.g., punctuation). 
    \item Convert all text to lowercase for case-insensitive matching. 
    \item Tokenize the cleaned description into small, well-formed tokens. These tokens make it easier to match codes to query fragments. 
  \end{enumerate}
  \item {Hierarchical Search (Top-Down)}:
  \begin{enumerate}
    \item Start at the top level of the ICD-10 hierarchy and move downward level by level.
    \item At each level:
    \begin{enumerate}
      \item If there are matches from the previous level, restrict the search to their child nodes.
      \item Otherwise, consider all descriptions at the current level.
      \item Apply the matching strategy:
      \begin{enumerate}
        \item Attempt exact token matches between query tokens and description tokens.
        \item If no exact matches are found, use approximate matching by calculating textual similarity using Levenshtein distance.
        \item If the similarity exceeds a predefined threshold, consider the description a potential match.
      \end{enumerate}
      \item If a better match is found at this level, update the best match and proceed to the next level.
      \item If no better match is found, terminate the hierarchical search.
    \end{enumerate}
  \end{enumerate}

  \item {Random Search (Flat Search)}:
  \begin{enumerate}
    \item Randomly search across all levels of the hierarchy without parent-child constraints. 
    \item For each node, apply the same matching strategy as above.
    \item Track the best match and score across all nodes visited.
    \item If the random search score is higher than the best hierarchical search score, use the new matched node as the starting point and consider all its descendants as matches.
  \end{enumerate}
    \item Return the ICD-10 codes, descriptions, and descendants corresponding to the highest match score. If no matches meet the threshold, indicate that no results were found for the user query.
\end{enumerate}


The hybrid approach enhances the algorithm's balance between precision and adaptability, accommodating both the hierarchical structure and the non-standard or semantically varying inputs.


\section{Experiments}
\label{sec:experiments}

We now evaluate the effectiveness of SearchAI in interpreting ICD-10 codes. Our goals are four-fold: (i)~evaluate the accuracy of SearchAI in interpreting ICD-10 codes from natural language, (ii)~analyze the robustness of our approach against different variations of the user questions, (iii)~measure and evaluate the query latencies with SearchAI and check whether it is interactive, and finally (iv)~test the scalability of SearchAI with the size of the dataset.

Below, we first describe the setup before discussing each of the experiments.


\subsection{Setup}

\paragraph*{Dataset.} We used two different datasets in our experiments. 
\begin{enumerate}
  \item \textit{FFS Dataset:} A dataset consisting of Medicare fee-for-service (FFS) claims data from Data.CMS.gov (2025). Each claim in this dataset includes up to 25 disease codes, as well as drug and other related codes. This dataset is used for evaluating SearchAI's performance in a controlled environment where the data structure is standardized and consistent.
  \item \textit{Production Dataset:} A dataset containing anonymized patient records linked to diseases, medical procedures, lab tests, and other clinical data. This dataset is used to evaluate SearchAI’s performance in handling large, complex data structures in real-world settings, where the data is more varied and unpredictable compared to fixed synthetic datasets like the FFS.  
\end{enumerate}

\paragraph*{Hardware.}
We ran all experiments on a MacOS 14.3.1 system with a 2.6 GHz 6-Core Intel Core i7 processor. SearchAI is implemented in Python, and the experiments were run in the PyCharm IDE. All external libraries, such as Levenshtein for similarity calculations, are pre-installed on the same machine.



\subsection{Accuracy}



\begin{figure}[!t]
  \centering
  \begin{subfigure}[t]{0.495\textwidth}
    \centering
    \includegraphics[width=\textwidth]{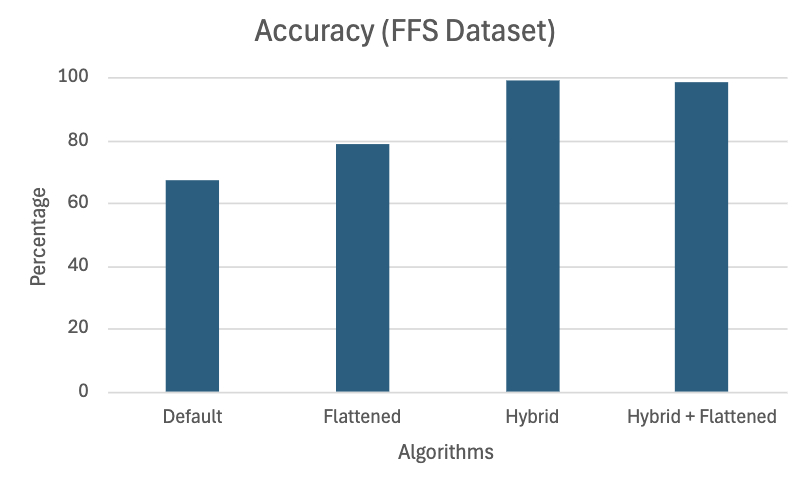}
    \caption{FFS dataset.}
    \label{fig:ffs-dataset-accuracy}
  \end{subfigure}
  \hfill
  \begin{subfigure}[t]{0.495\textwidth}
    \centering
    \includegraphics[width=\textwidth]{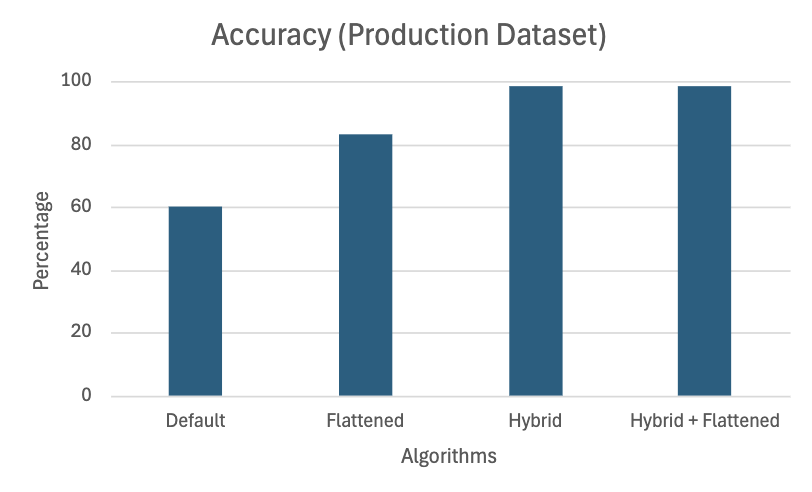}
    \caption{Production dataset.}
    \label{fig:production-dataset-accuracy}
  \end{subfigure}
  \caption{SearchAI accuracy on FFS and production datasets.}
  \label{fig:searchai-accuracy}
\end{figure}

We first evaluate the accuracy of SearchAI in identifying ICD-10 codes. Specifically, we measure accuracy as the overall proportion of cases where the SearchAI correctly retrieves the intended ICD-10 code. To do this, we convert each ICD code description into a corresponding question and assess how well the model interprets the information and retrieves all relevant ICD codes. 

Figures~\ref{fig:ffs-dataset-accuracy} and~\ref{fig:production-dataset-accuracy} show the results on FFS and production datasets respectively. 
The default version of hierarchical predictor has poor accuracies of $67.35\%$ and $60\%$ on FFS and production datasets respectively. However, the accuracy improves significantly with flattened and hybrid variants of the algorithm. For the FFS dataset, level-1 accuracies reached $99\%$ and $98.63\%$, while for the production dataset, the level-1 accuracies reached $98.3\%$ and $98.6\%$. This is a significant improvement of around $30\%$ increase compared to the initial accuracies of $60\%$ and $67.35\%$. 

Note that the accuracy of hybrid and flattened combined on the FFS dataset decreases slightly compared to the hybrid variant alone without any flattening. This is because the flattening process does not provide any additional accuracy on smaller FFS datasets with limited disambiguation needed. On the contrary, flattening FFS dataset makes it susceptible to missing valid branches.
Therefore, dataset size and hierarchy distribution are important factors for SearchAI. Adapting the design choices to dataset characteristics will be part of future work.


\subsection{Robustness}



\begin{figure}[!t]
  \centering
  \begin{subfigure}[t]{0.495\textwidth}
    \centering
    \includegraphics[width=\textwidth]{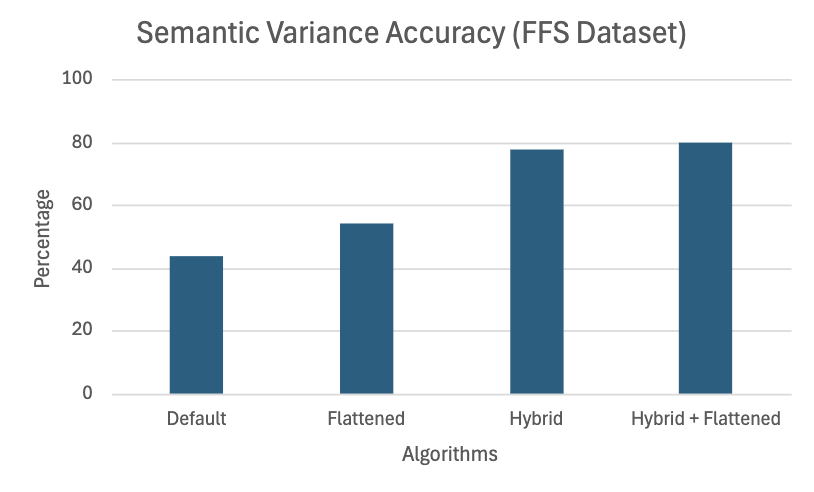}
    \caption{FFS dataset.}
    \label{fig:ffs-dataset-semantic-variance}
  \end{subfigure}
  \hfill
  \begin{subfigure}[t]{0.495\textwidth}
    \centering
    \includegraphics[width=\textwidth]{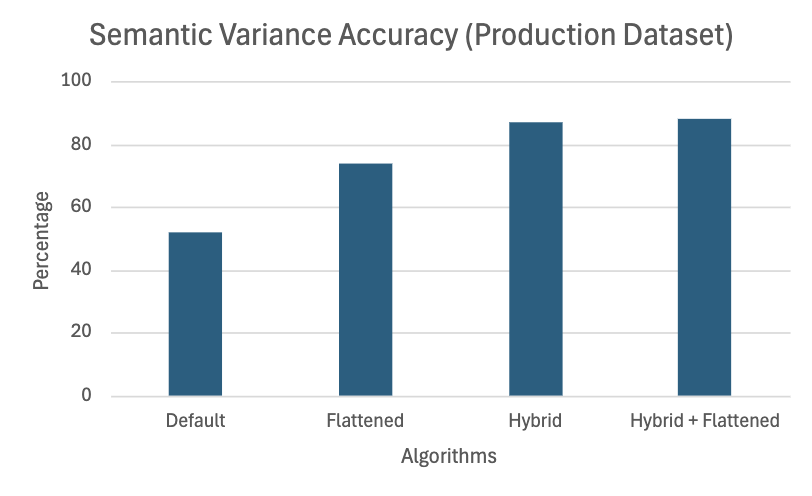}
    \caption{Production dataset.}
    \label{fig:production-dataset-semantic-variance}
  \end{subfigure}
  \caption{SearchAI accuracy with query variations on FFS and production datasets.}
  \label{fig:semantic-variance-accuracy}
\end{figure}

We now test the robustness of SearchAI. The idea is to evaluate how the accuracy of SearchAI varies as queries change.
To test robustness, we rephrased each code description in level-1 of both datasets using ChatGPT to produce semantically similar variants. We use the following prompt in ChatGPT to generate these variants of the original query: 

\vspace{0.2cm}
{\it ``For each test case, avoid using words from the original description and generate a semantically equivalent search query that asks how a physician would record the disease or health condition. Do not put the query in the form of a question. Return the modified test cases as the output.''}
\vspace{0.2cm}

Once we get the semantic variations of questions, we evaluate the accuracy for each of the algorithmic variants.
Figures~\ref{fig:ffs-dataset-semantic-variance} and~\ref{fig:production-dataset-semantic-variance} show the result for interpreting level-1 ICD-10 codes on FFS and production datasets, respectively. 
We see that robustness improves with flattened and hybrid variants.  
In fact, the combined hybrid and flattened variant achieved $79.86\%$ accuracy on the FFS dataset and $88.23\%$ on the production dataset, compared to the accuracies of $52\%$ and $43.88\%$ in other variants.
Note that we do expect semantic query variations to have lower accuracy rates, since the queries may change significantly compared to the original descriptions. Still, 
our goal is to reduce the degradation and ensure greater consistency and reliability of the results. 


\subsection{Latency}

\begin{figure}[!t]
  \centering
  \begin{subfigure}[t]{0.495\textwidth}
    \centering
    \includegraphics[width=\textwidth]{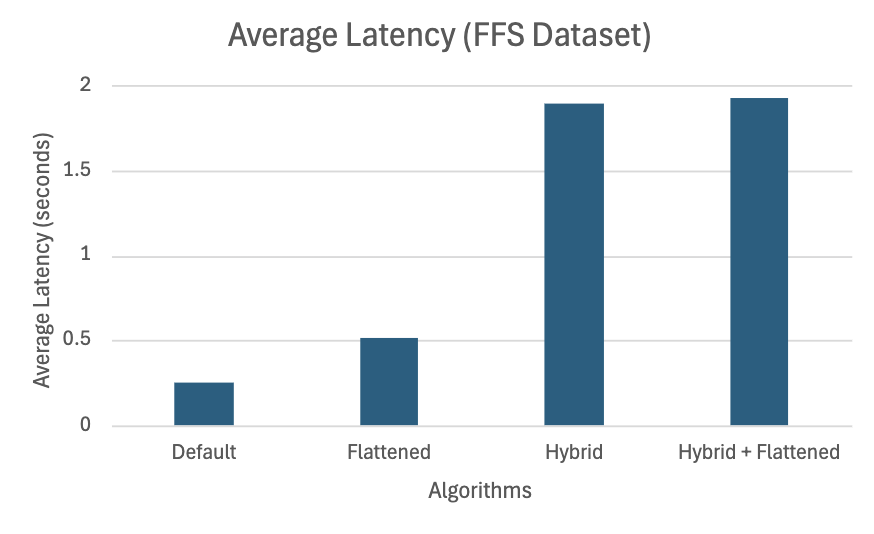}
    \caption{FFS dataset.}
    \label{fig:ffs-dataset-average-latency}
  \end{subfigure}
  \hfill
  \begin{subfigure}[t]{0.495\textwidth}
    \centering
    \includegraphics[width=\textwidth]{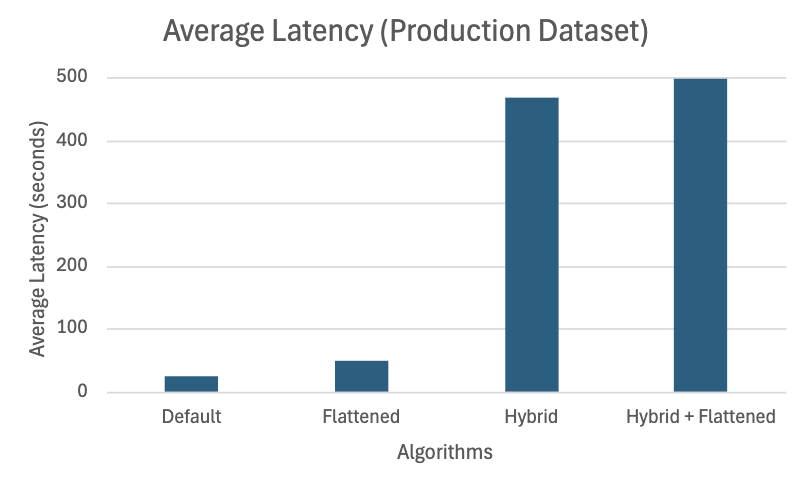}
    \caption{Production dataset.}
    \label{fig:production-dataset-average-latency}
  \end{subfigure}
  \caption{Average latency on FFS and production datasets.}
  \label{fig:average-latency}
\end{figure}

We now measure the query latencies in SearchAI. 
We record latency as the time SearchAI takes to compute all ICD Codes for all queries in level-1. We ran the latency experiment for each dataset and for each algorithm variant. We used the {\it time} package in Python to evaluate the time taken.
Figures~\ref{fig:ffs-dataset-average-latency} and~\ref{fig:production-dataset-average-latency} show the average latency of SearchAI in interpreting level-1 ICD-10 codes on FFS and production datasets, respectively. 

  
The above results show reasonable latencies with SearchAI (note that they are for as many queries as the number of codes in each dataset), though there are notable differences between different variants of the hierarchical predictor. 
Looking at specific latency numbers to answer all level-1 queries,
the default hierarchical predictor performs the fastest on both datasets, with the production dataset averaging $26.60$ seconds for looking up all $4{,}782$ ICD codes.
  
The more sophisticated hybrid variant and combined hybrid and flattened variant exhibit significantly longer latencies, with the production dataset taking over $8$ minutes for looking up all $4{,}782$ ICD codes (total time for as many queries as the number of codes in the dataset). 
This is because the hybrid variant performs more expensive random search to find a good starting point first. This initial search is purely a meta operation without doing any actual match.

\subsection{Scalability}

\begin{figure}[!t]
  \centering
  \begin{subfigure}[t]{0.495\textwidth}
    \centering
    \includegraphics[width=\textwidth]{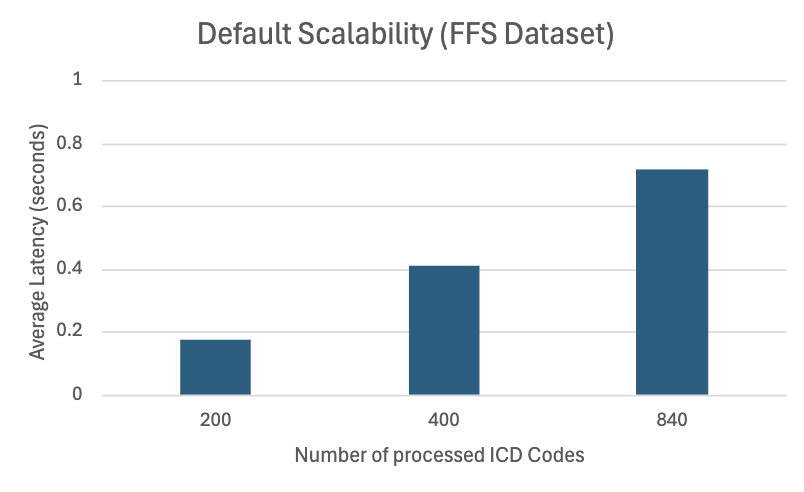}
    \caption{Default variant.}
    \label{fig:ffs-default-scalability}
  \end{subfigure}
  \hfill
  \begin{subfigure}[t]{0.495\textwidth}
    \centering
    \includegraphics[width=\textwidth]{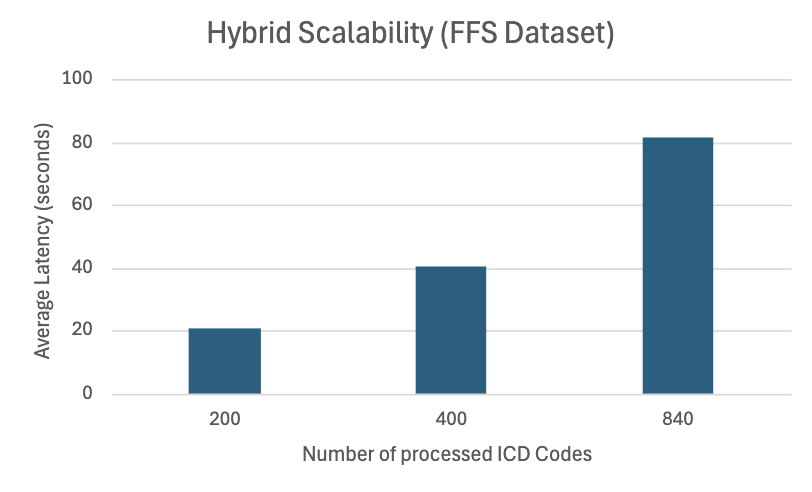}
    \caption{Hybrid variant.}
    \label{fig:ffs-hybrid-scalability}
  \end{subfigure}
  \caption{Scalability using default and hybrid variant on FFS dataset.}
  \label{fig:ffs-scalability}
\end{figure}


We now evaluate scalability by varying the number of ICD codes in the dataset and measuring the time taken to answer all queries on those modified datasets.
Specifically, we break down the two datasets in our experiments, namely the FFS dataset and the production dataset, into three fractional ones: $25\%$ of the codes, $50\%$ of the codes, and $100\%$ of the entire dataset. 
We do not consider flattening in this experiment since it is applicable only to ICD codes for levels 0 and 1, and fractional datasets may not cover those levels fully.

\begin{figure}[!t]
  \centering
  \begin{subfigure}[t]{0.495\textwidth}
    \centering
    \includegraphics[width=\textwidth]{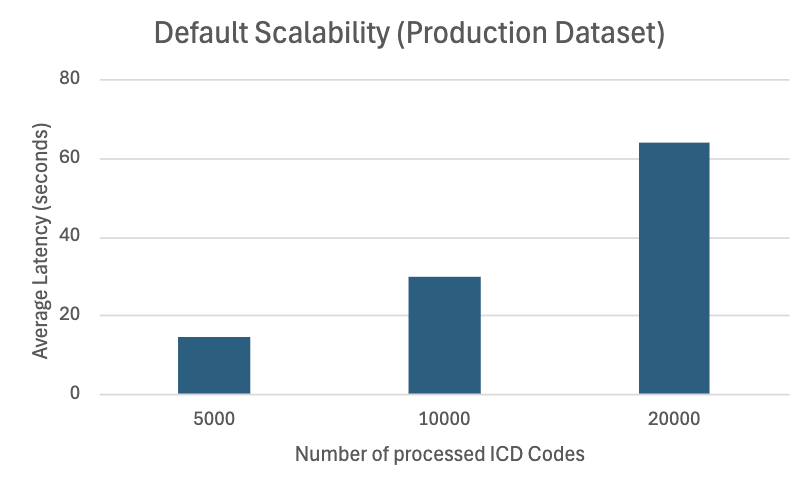}
    \caption{Default variant.}
    \label{fig:prod-default-scalability}
  \end{subfigure}
  \hfill
  \begin{subfigure}[t]{0.495\textwidth}
    \centering
    \includegraphics[width=\textwidth]{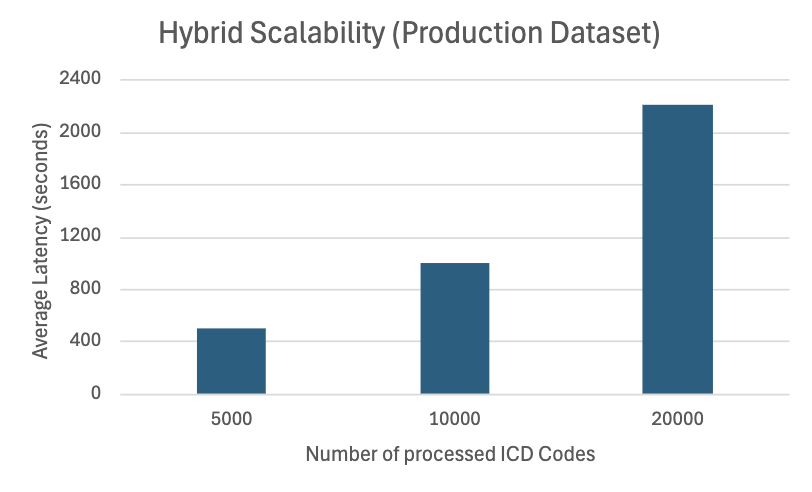}
    \caption{Hybrid variant.}
    \label{fig:prod-hybrid-scalability}
  \end{subfigure}
  \caption{Scalability using default and hybrid variant on production dataset.}
  \label{fig:prod-scalability}
\end{figure}

Figures~\ref{fig:ffs-default-scalability}  show the average latency of SearchAI's default variant in interpreting all codes on FFS dataset (sizes $200$, $400$, and $840$).
Figure~\ref{fig:ffs-hybrid-scalability}
shows the same result on the hybrid variant of hierarchical predictor.
In both cases, we see that SearchAI latencies scales linearly with the dataset sizes.
Figures~\ref{fig:prod-default-scalability} and~\ref{fig:prod-hybrid-scalability} show the
scalability results for production dataset. The fractional dataset sizes in this case are $5{,}000$, $10{,}000$, and $20{,}000$, and again we show results on both the default and hybrid variants.

The scalability results show SearchAI's performance with varying dataset sizes, and both datasets displayed good scaling. As the number of codes increases, the query latencies increase linearly. However, as noted earlier, more complex algorithms such as the hybrid variant are challenging with larger datasets, and it takes around $40$ minutes to compute all $20{,}000$ ICD codes in the production dataset. Of course, no single query can realistically ask for that many codes, and so this is not the response time that users will see. The per-query response time for a single code is just $110$ milliseconds in the worst case.
Future work will investigate more aggressive computation sharing when user queries retrieve a large number of codes.

\vspace{0.2cm}
In summary, the experimental results show that SearchAI can answer natural language queries with very high accuracy that are robust even when the questions change (while keeping the same semantic meaning). SearchAI also has low latency (from a few milliseconds to a hundred milliseconds in the worst case) and scales well with dataset sizes.

\section{Conclusion}
Healthcare practitioners often have complex search queries, requiring a clear understanding and accurate mapping to relevant medical codes.
In this paper, we presented a generative AI approach, SearchAI, to help search clinical data with high accuracy. 
Unlike conventional code assignment, which maps a given condition to a single code, clinical data search is a one-to-many problem requiring accurate hierarchical lookup.

SearchAI interprets natural language queries into a family of medical codes and other related filters and generates detailed descriptive answers.
SearchAI is more than $98\%$ accurate and outperforms the manual search approach in terms of accuracy, robustness, performance, and scalability. 
As a result, SearchAI can improve healthcare workflows, reduce administrative burdens, and enhance both patient and provider outcomes.

\section*{Acknowledgments}
This paper was formatted using the \emph{Arxiv \& PRIME AI Style Template} by Moulay A. Akhloufi \cite{overleaf-template}, available on Overleaf under a CC BY 4.0 license.

\bibliographystyle{unsrt}  
\bibliography{references}

\end{document}